\documentclass[english,aip,pop,showpacs,preprint,superscriptaddress]{revtex4-1}
\usepackage{lmodern}
\usepackage{lmodern}
\usepackage[T1]{fontenc}
\usepackage[latin9]{inputenc}
\usepackage{amsmath}
\usepackage{amssymb}
\usepackage{esint}

\makeatletter

\usepackage{amsthm}\usepackage{latexsym}\usepackage{bm}\usepackage{amsfonts}

\usepackage{babel}

\usepackage{babel}

\usepackage{hyperref}

\makeatother

\usepackage{babel}
\begin{document}

\title{A lattice Maxwell system with discrete space-time symmetry and local
energy-momentum conservation}

\author{Jianyuan Xiao}

\affiliation{Department of Modern Physics, University of Science and Technology
of China, Hefei, Anhui 230026, China}

\author{Hong Qin }
\email{hongqin@ustc.edu.cn}

\affiliation{Department of Modern Physics, University of Science and Technology
of China, Hefei, Anhui 230026, China}

\affiliation{Plasma Physics Laboratory, Princeton University, Princeton, NJ 08543,
U.S.A}

\author{Yuan Shi }

\affiliation{Plasma Physics Laboratory, Princeton University, Princeton, NJ 08543,
U.S.A}

\author{Jian Liu}

\affiliation{Department of Modern Physics, University of Science and Technology
of China, Hefei, Anhui 230026, China}

\author{Ruili Zhang}

\affiliation{School of Science, Beijing Jiaotong University, Beijing 100044, P.R. China }
\begin{abstract}
A lattice Maxwell system is developed with gauge-symmetry, symplectic
structure and discrete space-time symmetry. Noether's theorem for
Lie group symmetries is generalized to discrete symmetries for the
lattice Maxwell system. As a result, the lattice Maxwell system is
shown to admit a discrete local energy-momentum conservation law corresponding
to the discrete space-time symmetry. These conservative properties
make the discrete system an effective algorithm for numerically solving
the governing differential equations on continuous space-time. Moreover,
the lattice model, respecting all conservation laws and geometric
structures, is as good as and probably more preferable than the continuous
Maxwell model. Under the simulation hypothesis by Bostrom \cite{Bostrom03}
and in consistent with the discussion on lattice QCD by Beane et al.
\cite{Beane12}, the two interpretations of physics laws on space-time
lattice could be essentially the same. 
\end{abstract}



\maketitle
\global\long\def\EXP{\times10}
 \global\long\def\rmd{\mathrm{d}}
 \global\long\def\diag{\textrm{diag}}
 \global\long\def\xs{ \mathbf{x}_{s}}
 \global\long\def\bfx{\mathbf{x}}
 \global\long\def\bfv{\mathbf{v}}
 \global\long\def\bfA{\mathbf{A}}
 \global\long\def\bfB{\mathbf{B}}
 \global\long\def\bfS{\mathbf{S}}
 \global\long\def\bfG{\mathbf{G}}
 \global\long\def\bfE{\mathbf{E}}
 \global\long\def\bfM{\mathbf{M}}
 \global\long\def\bfQ{\mathbf{Q}}
 \global\long\def\bfu{\mathbf{u}}
 \global\long\def\bfe{\mathbf{e}}
 \global\long\def\bfd{\mathbf{d}}
 \global\long\def\rme{\mathrm{e}}
 \global\long\def\rmi{\mathrm{i}}
 \global\long\def\rmq{\mathrm{q}}
 \global\long\def\ope{\omega_{pe}}
 \global\long\def\oce{\omega_{ce}}
 \global\long\def\FIG#1{Fig.~#1}
 \global\long\def\EQ#1{Eq.~(\ref{#1})}  \global\long\def\SEC#1{Sec.~#1}
\global\long\def\REF#1{Ref.~\cite{#1}}
 \global\long\def\DDELTAT#1{\textrm{Dt}\left(#1\right)}
 \global\long\def\DDELTATA#1{\textrm{Dt}^{*}\left(#1\right)}
 \global\long\def\CURLD{ {\mathrm{curl_{d}}}}
 \global\long\def\DIVD{ {\mathrm{div_{d}}}}
 \global\long\def\CURLDP{ {\mathrm{curl_{d}}^{T}}}
 \global\long\def\cpt{\captionsetup{justification=raggedright }}
 \global\long\def\act{\mathcal{A}}
 \global\long\def\calL{\mathcal{L}}
 \global\long\def\calJ{\mathcal{J}}
 \global\long\def\DELTAA{\left( \bfA_{J,l}-\bfA_{J,l}' \right)}
 \global\long\def\DELTAAL{\left( \bfA_{J,l-1}-\bfA_{J,l-1}' \right)}
 \global\long\def\ADAGGER{\bfA_{J,l}^{\dagger}}
 \global\long\def\ADAGGERA#1{\bfA_{J,#1}^{x/2}}
 \global\long\def\EDAGGER#1{\bfE_{J,#1}^{x/2}}
 \global\long\def\BDAGGER#1{\bfB_{J,#1}^{x/2}}
 \global\long\def\DDT{\frac{\partial}{\partial t}}
 \global\long\def\DBYDT{\frac{\rmd}{\rmd t}}
 \global\long\def\DBYANY#1{\frac{\partial}{\partial#1}}
 \global\long\def\WZERO#1{W_{\sigma_{0} I}\left( #1 \right)}
 \global\long\def\WONE#1{W_{\sigma_{1} J}\left( #1 \right)}
 \global\long\def\WONEJp#1{W_{\sigma_{1} J'}\left( #1 \right)}
 \global\long\def\WTWO#1{W_{\sigma_{2} K}\left( #1 \right)}
 \global\long\def\MQQ{M_{00}}
 \global\long\def\MDQDQ{M_{11}}
 \global\long\def\MDQQ{M_{01}}

Maxwell's equations are partial differential equations (PDEs) governing
the dynamics of electromagnetic field on space-time. In this paper,
we propose a lattice Maxwell system with gauge-symmetry, symplectic
structure, discrete space-time symmetry and discrete local energy-momentum
conservation. The electromagnetic field is only defined on a space-time
lattice with a set of discrete rules governing its dynamics. A lattice
Maxwell system can be interpreted in two ways. First, it can be viewed
as a model for the electromagnetic field by itself. Second, it can
be treated as a numerical algorithm of the continuous Maxwell's equations.
For both purposes, it is desirable for the lattice Maxwell system
to conserve fundamental physics quantities locally, such as energy-momentum,
and preserve important structures, such as space-time symmetry, gauge-symmetry
and symplectic structure. Especially for the first purpose, the conservation
and structure preserving properties are indispensable. One can argue
that a lattice Maxwell system which respects all the important conservation
laws and structures is a valid model for the electromagnetic field.
It is as good as the continuous Maxwell's equations, but with an added
advantage that it can be easily calculated. For the second purpose,
i.e., using the lattice Maxwell system as an algorithm, the conservative
and structure-preserving properties of the algorithm, will render
more reliable numerical solutions of the continuous Maxwell's equations.
This is especially true for simulations of long-term multi-scale dynamics.
A well-known example in this respect is Yee's algorithm \cite{yee1966numerical}
and its generalizations using discrete exterior calculus (DEC). The
outstanding performance of Yee's algorithm is attributed to the fact
that it preserves the differential form structure of the Maxwell system
\cite{desbrun2005discrete,stern2007geometric,desbrun2008discrete}.
We emphasize that a local energy-momentum conservation law for numerical
algorithms is a much more desirable property than global energy-momentum
conservation law. This is because a local conservation law requires
the numerical solutions satisfy the law at every space-time grid point,
and the number of constraints is as many as the space-time grid points.
For a global conservation law, there is only one conserved quantity,
and this is not a strong constraint for the system which contains
a large number of degrees of freedom defined on the space-time lattice. 

Under the simulation hypothesis by Bostrom \cite{Bostrom03} and in
consistent with the discussion on lattice QCD by Beane et al. \cite{Beane12},
the two interpretations of physics laws on space-time lattice could
be essentially the same. 

How to design a lattice Maxwell system with these desirable conservative
and structure-preserving properties? It is natural and probably necessary
to adopt a field theoretical approach. In this paper, the lattice
Maxwell system with gauge symmetry, symplectic structure, discrete
space-time symmetry and discrete local energy-momentum conservation
is constructed by discretizing the Lagrangian of the electromagnetic
field on a space-time lattice. 

For continuous systems, Noether's theorem \cite{noether1971invariant}
connects Lie group symmetries and conservation laws. For example,
energy-momentum conservation is the consequence of space-time symmetry.
However, for discrete systems, such a connection has not been fully
understood. Even though Noether's theorem has been applied to discrete
systems \cite{lee1987difference,wendlandt1997mechanical,marsden2001discrete,hairer2006geometric,Hydon14},
the symmetries in these applications are continuous Lie groups, as
required by the Noether's theorem. In this paper, we demonstrate a
generalized version of Noether's theorem which establishes connections
between discrete symmetries and discrete conservation laws for the
lattice Maxwell system. As far as we know, such a generalized Noether's
theorem for discrete symmetries has not be discussed in the existing
literature. We then show that the lattice Maxwell system admits a
discrete space-time symmetry, and more importantly that the discrete
space-time symmetry induces a discrete local energy-momentum conservation
law.

We should mention that there are other methods to construct numerical
schemes with conservation properties for partial differential equations,
e.g., the finite volume method and the discontinuous Galerkin method.
In particular, Poynting's theorem for Yee's algorithm has been discussed
\cite{chew1994electromagnetic,de1995poynting}. However, these studies
are only on the level of numerical algorithms, and are not related
to the fundamental properties of discrete space-time symmetries. 

The lattice Maxwell system also admits a gauge symmetry and conserves
a finite dimensional symplectic structure. As in the continuous case,
the gauge symmetry on the lattice generates a discrete charge conservation
law \cite{squire2012geometric,xiao2015explicit}. The conservation
of the symplectic structure inherited from the variational structure
guarantees the conservation of phase-space volume and bounds long-term
errors in energy \cite{yee1966numerical,stern2007geometric,squire2012geometric,xiao2013variational,xiao2015explicit,xiao2015variational}.
Since these topics have been well studied, they will not be the main
focus of the present paper. 

We start our discussion from the action of the electromagnetic field
on the continuous space-time,
\begin{eqnarray}
\act[\bfA] & = & \int\rmd t\rmd\bfx\calL\left(\bfA,\phi\right)~,\\
\calL\left(\bfA,\phi\right) & = & \frac{1}{2}\left(\left(-\nabla\phi-\dot{\bfA}\right)^{2}-\left(\nabla\times\bfA\right)^{2}\right)~,
\end{eqnarray}
where $\calL$ is the Lagrangian density. For simplicity, permeability
$\mu_{0}$ and permittivity $\varepsilon_{0}$ are set to 1. The Euler-Lagrange
equations for $\bfA,\phi$ are
\begin{eqnarray}
\frac{\partial}{\partial t}\left({\dot{\bfA}+\nabla\phi}\right)-\nabla\times\nabla\times\bfA & = & 0~,\\
\nabla\cdot{\left(\dot{\bfA}+\nabla\phi\right)} & = & 0~.
\end{eqnarray}
We select a ``cubic'' lattice in space-time. The discrete Lagrangian
density is chosen to be
\begin{eqnarray}
\calL_{J,l} & = & \frac{1}{2}\left(\DDELTAT{\bfA_{J,l-1}}-\nabla_{d}\phi_{J,l}\right)^{2}-\frac{1}{2}\left(\CURLD\bfA_{J,l}\right)^{2}~,\\
\DDELTAT{f_{l}} & \equiv & \frac{f_{l+1}-f_{l}}{\Delta t}~,
\end{eqnarray}
where $J$ is the spatial grid indices $[i,j,k]$, and $l$ is the
temporal index, and $\nabla_{d}$ and $\CURLD$ are discrete gradient
and discrete curl operators, respectively. They are defined by Eqs.\,\eqref{EqnDEFGRADD}and
\eqref{EqnDEFCURLD} in the Appendix. The discrete action is the sum
of the discrete Lagrangian density,
\begin{eqnarray}
\act_{d}[\bfA,\phi] & = & \sum_{J,l}\calL_{J,l}\Delta t\Delta x^{3}~.
\end{eqnarray}
The Lagrangian density $\calL_{J,l}$ admits a gauge symmetry, meaning
that the following transformation 
\begin{eqnarray}
\bfA_{J,l} & \rightarrow & \bfA_{J,l}+\nabla_{d}\psi_{J,l}~,\label{eq:gauge-symA}\\
\phi_{J,l} & \rightarrow & \phi_{J,l}-\DDELTATA{\psi_{J,l}}~,\label{eq:gauge-symphi}
\end{eqnarray}
will leave $\calL_{J,l}$ invariant. Here, Dt$^{*}$ is the dual operator
of of Dt, 
\[
\DDELTATA{f_{l}}\equiv\frac{f_{l}-f_{l-1}}{\Delta t}~.
\]
We note that this symmetry is defined on the space-time lattice. However,
the symmetry group itself is continuous and forms a Lie group. Due
to the gauge symmetry, we can choose a discrete gauge to simplify
the calculation. In the present study, the temporal gauge is adopted,
i.e., $\phi_{J,l}=0$. In this gauge, the discrete Euler-Lagrange
equation for $\bfA_{J,l}$ is\cite{marsden2001discrete} 
\begin{eqnarray}
\DDELTATA{\DDELTAT{\bfA_{J,l}}}+\CURLDP\CURLD\bfA_{J,l} & = & 0~.\label{EqnDEL}
\end{eqnarray}

Now we introduce the concept of discrete symmetry. A discrete transformation
of the lattice field $\bfA_{J,l}$, 
\begin{eqnarray}
\bfA_{J,l}\rightarrow\bfA_{J,l}'~,\label{EqnDTR}
\end{eqnarray}
is a discrete symmetry, if the resulting variation of the Lagrangian
density is at most a discrete 4-divergence of a discrete 4-vector
field $(\calL_{t,J,l},\calL_{\bfx,J,l})$, i.e.,
\begin{eqnarray}
\Delta\calL_{J,l}=\calL_{J,l}(\bfA_{J,l})-\calL_{J,l}(\bfA_{J,l}')=\DDELTATA{\calL_{t,J,l}}+\DIVD^{*}{\calL_{\bfx,J,l}}~.
\end{eqnarray}
Here, $\calL_{t,J,l}$ and $\calL_{\bfx,J,l}$ are the temporal and
spatial components of the 4-vector field, which only depends on the
values of discrete vector potential $\bfA$ near the grid point $J,l$,
e.g., $\bfA_{J,l},\bfA_{J,l+1},\bfA_{i,j-1,k,l},\dots~.$ The operator
$\DIVD^{*}$ is defined in \EQ{EqnDEFDIVD1}. This definition of
discrete symmetry is similar to that of continuous (Lie group) symmetry
in Noether's theorem \cite{olver1993applications331}. We emphasize
again that the gauge symmetry of Eqs.\,\eqref{eq:gauge-symA} and
\eqref{eq:gauge-symphi} is a continuous Lie group symmetry, instead
of a discrete symmetry. We now show that for the lattice Maxwell system
proposed here, a discrete symmetry will induce a discrete local conservation
law. This result can be viewed as a discrete generalization of Noether's
theorem. However, the fundamental difference is that the symmetry
group in the current context is a discrete group, instead of a Lie
group for the standard Noether's theorem. 

To prove this fact, we first calculate the change of the Lagrangian
density due to the transformation \eqref{EqnDTR},
\begin{eqnarray}
\Delta\calL_{J,l} & = & \frac{1}{2}\DDELTAT{\bfA_{J,l-1}}^{2}-\frac{1}{2}\left(\CURLD\bfA_{J,l}\right)^{2}-\frac{1}{2}\DDELTAT{\bfA_{J,l-1}'}^{2}+\frac{1}{2}\left(\CURLD\bfA_{J,l}'\right)^{2}\\
 & = & \DDELTAT{\bfA_{J,l-1}^{\dagger}}\cdot\DDELTAT{\DELTAAL}-\CURLD\bfA_{J,l}^{\dagger}\cdot\CURLD\DELTAA~,\label{EqnDELTAL}
\end{eqnarray}
where
\begin{eqnarray}
\ADAGGER\equiv\frac{1}{2}\left(\bfA_{J,l}+\bfA_{J,l}'\right)~.
\end{eqnarray}
Note that $\Delta\calL_{J,l}$ cannot be made arbitrarily small as
in the case of Lie group symmetries in the standard Noether's theorem.
Similar to Eq.\,\eqref{EqnDEL}, the Euler-Lagrange equation for
$\bfA_{J,l}'$ derived from $\calL_{J,l}(\bfA_{J,l}')$ is 

\begin{eqnarray}
\DDELTATA{\DDELTAT{\bfA_{J,l}'}}+\CURLDP\CURLD\bfA_{J,l}' & = & 0~.\label{EqnDELP}
\end{eqnarray}
The combination of Eq.\,$(\ref{EqnDELTAL})+(\textrm{Eq}.\thinspace(\ref{EqnDEL})+\textrm{Eq.\thinspace}(\ref{EqnDELP}))\cdot\frac{1}{2}\left(\bfA_{J,l}-\bfA_{J,l}'\right)$
gives 
\begin{eqnarray}
\Delta\calL_{J,l}~= & \DDELTAT{\bfA_{J,l-1}^{\dagger}}\cdot\DDELTAT{\DELTAAL}-\CURLD\bfA_{J,l}^{\dagger}\cdot\CURLD\DELTAA\nonumber \\
 & +\DDELTATA{\DDELTAT{\bfA_{J,l}^{\dagger}}}\cdot\left(\bfA_{J,l}-\bfA_{J,l}'\right)+\left(\CURLDP\CURLD\bfA_{J,l}^{\dagger}\right)\cdot\left(\bfA_{J,l}-\bfA_{J,l}'\right)\\
= & \DDELTATA{\DDELTAT{\ADAGGER}\cdot\DELTAA}-\DIVD^{*}\left(\DELTAA*\times\CURLD\ADAGGER\right)\label{EqnDDLDDT}
\end{eqnarray}
In deriving \EQ{EqnDDLDDT}, use has been made of the operator identity
\EQ{EqnCURLCUR2LDIV} in the Appendix. Now, if the discrete transformation
$\bfA_{J,l}\rightarrow\bfA_{J,l}'$ is a discrete symmetry, then by
definition $\Delta\calL_{J,l}$ is the discrete 4-divergence of a
discrete 4-vector $(\calL_{t,J,l},\calL_{\bfx,J,l})$,
\begin{eqnarray}
\Delta\calL_{J,l} & = & \DDELTATA{\calL_{t,J,l}}+\DIVD^{*}\calL_{\bfx,J,l}~.\label{eq:dljl}
\end{eqnarray}
Then, Eqs.\,\eqref{EqnDDLDDT} and \eqref{eq:dljl} can be combined
to give a discrete local conservation law,
\begin{eqnarray}
\DDELTATA{\DDELTAT{\bfA_{J,l}^{\dagger}}\cdot\DELTAA-\calL_{t,J,l}}-\nonumber \\
\DIVD^{*}\left(\DELTAA*\times\CURLD\ADAGGER+\calL_{\bfx,J,l}\right) & = & 0~.\label{EqnDNT0}
\end{eqnarray}
This completes the proof of the generalized Noether's theorem for
discrete symmetries of the lattice Maxwell system.

In the present study, we will consider the following spatial and the
temporal translation symmetries, 
\begin{eqnarray}
\bfA_{J,l}\rightarrow\bfA_{J-x,l}~,\label{EqnAOFFX}\\
\bfA_{J,l}\rightarrow\bfA_{J-y,l}~,\\
\bfA_{J,l}\rightarrow\bfA_{J-z,l}~,\\
\bfA_{J,l}\rightarrow\bfA_{J,l+1}~,\label{EqnAOFFT}
\end{eqnarray}
where $J-x$, $J-y$, $J-z$ denote $[i-1,j,k]$, $[i,j-1,k]$, $[i,j,k-1]$,
respectively. First, let's look at the symmetry in the \EQ{EqnAOFFX},
which is the discrete translation symmetry in the $\bfe_{x}$-direction.
In this case, the change of Lagrangian density is 
\begin{eqnarray}
\Delta\calL_{J,l} & = & \frac{1}{2}\left(\DDELTAT{\bfA_{J,l-1}}^{2}-\DDELTAT{\bfA_{J-x,l-1}}^{2}-\left(\CURLD\bfA_{J,l}\right)^{2}+\left(\CURLD\bfA_{J-x,l}\right)^{2}\right)~,
\end{eqnarray}
which is the discrete 4-divergence of the following discrete 4-vector
\begin{eqnarray}
\calL_{t,J,l} & = & 0~,\\
\calL_{\bfx,J,l} & = & [\frac{1}{2}\DDELTAT{\bfA_{J,l-1}}^{2}-\frac{1}{2}\left(\CURLD\bfA_{J,l}\right)^{2},0,0]~.
\end{eqnarray}
This verifies that \EQ{EqnAOFFX} is indeed a discrete symmetry.
Therefore, according to \EQ{EqnDNT0}, we obtain the following discrete
conservation law, 
\begin{eqnarray}
\DDELTATA{\DDELTAT{\bfA_{J,l}^{x/2}}\cdot\left(\bfA_{J,l}-\bfA_{J-x,l}\right)}-\nonumber \\
\DIVD^{*}\left(\left(\bfA_{J,l}-\bfA_{J-x,l}\right)*\times\CURLD\bfA_{J,l}^{x/2}+[\frac{1}{2}\DDELTAT{\bfA_{J,l-1}}^{2}-\frac{1}{2}\left(\CURLD\bfA_{J,l}\right)^{2},0,0]\right) & = & 0~.\label{EqnDNT1}
\end{eqnarray}
Here, 
\begin{equation}
\bfA_{J,l}^{x/2}\equiv\frac{\bfA_{J,l}+\bfA_{J-x,l}}{2}\thinspace.
\end{equation}
Equation \eqref{EqnDNT1} is the discrete local momentum conservation
law in the $\bfe_{x}$-direction. It can be transformed into a familiar
form expressed in terms of $\boldsymbol{\mathbf{E}}$ and $\mathbf{\boldsymbol{B}}$.
Taking a discrete divergence $\DIVD^{*}$ of \EQ{EqnDEL} and using
the discrete operator identity \EQ{EqnDIVDPCURLDP}, we can see
that 
\begin{eqnarray}
\DDELTATA{\DIVD^{*}\DDELTAT{\bfA_{J,l}}} & = & 0~.
\end{eqnarray}
If initially $\DIVD^{*}\DDELTAT{\bfA_{J,0}}$ is zero, which will
be automatically satisfied when there is no charge in the space, then
it remains zero all time. Therefore,
\begin{eqnarray}
\DIVD^{*}\left(\DDELTAT{\bfA_{J,l}}\right) & = & 0~,\label{eq:divd0}
\end{eqnarray}
The electromagnetic fields on the lattice are defined as 
\begin{eqnarray}
\bfE_{J,l} & = & -\DDELTAT{\bfA_{J,l}}~,\\
\bfB_{J,l} & = & (\CURLD\bfA)_{J,l}~.
\end{eqnarray}
 In terms of $\bfE_{J,l}$ and $\bfB_{J,l}$, the discrete Maxwell
equations are
\begin{eqnarray}
\DDELTATA{\bfE_{J,l}}=(\CURLDP\bfB)_{J,l}~,\label{EqnDMEE}\\
\DDELTAT{\bfB_{J,l}}=-(\CURLD\bfE)_{J,l}~.\label{EqnDMEB}
\end{eqnarray}

With the help of Eq.\,\eqref{eq:divd0} and discrete operator identities
(\ref{EqnCURLXDOTDOT}) and (\ref{EqnDIVDDOTDIRMUL}), \EQ{EqnDNT1}
can be transformed into 
\begin{eqnarray}
\DDELTATA{\bfE_{J,l}^{x/2}\times*\bfB_{J,l}}_{x}+\DIVD^{*}\left(\DDELTATA{\EDAGGER l}\otimes A_{x,J,l}-\left(\EDAGGER{l-1}\otimes E_{x,J,l-1}\right)\right)-\nonumber \\
\DIVD^{*}\left(-\left(\bfA_{J,l}-\bfA_{J-x,l}\right)*\times\CURLD\ADAGGERA l-[\frac{1}{2}\bfE_{J,l-1}^{2}-\frac{1}{2}\left(\CURLD\bfA_{J,l}\right)^{2},0,0]\right)_{x} & = & 0~.\label{EqnDNT3}
\end{eqnarray}
In deriving Eq.\,\eqref{EqnDNT3}, use is also made of 
\begin{equation}
\DDELTATA{\DIVD^{*}\left(\bfE_{J,l}^{x/2}\otimes A_{x,J,l}\right)}=\DIVD^{*}\left(\DDELTATA{\EDAGGER l}\otimes A_{x,J,l}-\left(\EDAGGER{l-1}\otimes E_{x,J,l-1}\right)\right).
\end{equation}
The second term in \EQ{EqnDNT3} can be simplified by identities
(\ref{EqnOTOT2CT2}) and (\ref{EqnDIVDPCURLDP}), 
\begin{eqnarray}
\DIVD^{*}\left(\DDELTATA{\EDAGGER l}\otimes A_{x,J,l}\right)=\DIVD^{*}\left(-\nabla_{d}A_{x,J,l}\times_{2}\BDAGGER l\right)~.
\end{eqnarray}
Finally, using identity \eqref{EqnOTXDOTCTXCT2}, the discrete local
momentum conservation law in the $\bfe_{x}$-direction in terms of
$\bfE_{J,l}$ and $\bfB_{J,l}$ is
\begin{eqnarray}
\DDELTATA{\EDAGGER l\times*\bfB_{J,l}}_{x}+\DIVD^{*}[\frac{1}{2}\bfE_{J,l-1}\cdot{\bfE_{J,l-1}},0,0]-\DIVD^{*}\left(\bfE^{x/2}\otimes E_{x}\right)_{J,l-1} & +\nonumber \\
\DIVD^{*}[\frac{\bfB_{J-x,l}\cdot\bfB_{J,l}}{2},0,0]-\DIVD^{*}\left(\bfB\otimes^{*}B_{x}^{x/2}\right)_{J,l} & = & 0~.
\end{eqnarray}
It is straightforward to verify that it recovers the familiar momentum
conservation law in the continuous space-time when the grid-size goes
to zero. The discrete local momentum conservation law in the $\bfe_{y}$
or $\bfe_{z}$ direction can be obtained in a similar way.

Next, we look at the discrete local energy conservation due to the
discrete temporal symmetry specified by \EQ{EqnAOFFT}. In this
case, the change of the Lagrangian density in terms of $\bfE_{J,l}$
and $\bfB_{J,l}$ is 
\begin{eqnarray}
\Delta\calL_{J,l} & = & \frac{1}{2}\left(\bfE_{J,l-1}^{2}-\bfE_{J,l-2}^{2}-\bfB_{J,l}^{2}+\bfB_{J,l-1}^{2}\right)~,
\end{eqnarray}
and it is the discrete 4-divergence of the discrete 4-vector field
\begin{eqnarray}
\calL_{t,J,l} & = & \frac{1}{2}\left(\bfE_{J,l-1}^{2}-\bfB_{J,l}^{2}\right)~,\\
\calL_{\bfx,J,l} & = & [0,0,0]~.
\end{eqnarray}
This verifies that $\bfA_{J,l}\rightarrow\bfA_{J,l+1}~$ is a discrete
symmetry. According to \EQ{EqnDNT0}, the local discrete conservation
law in terms of $\bfE_{J,l}$ and $\bfB_{J,l}$ for this discrete
symmetry is 
\begin{eqnarray}
\DDELTATA{\frac{1}{2}\bfE_{J,l}\cdot\bfE_{J,l-1}+\frac{1}{2}\bfB_{J,l}^{2}}+\DIVD^{*}\left(\bfE_{J,l-1}*\times\frac{\bfB_{J,l}+\bfB_{J,l-1}}{2}\right) & = & 0~.
\end{eqnarray}
Of course, this is the discrete local energy conservation law. It
recovers the well-known continuous local energy conservation law,
a.k.a. Ponyting's theorem, when the grid-size approaches zero. 

In conclusion, we reported three important advances in the study of
lattice model and structure-preserving geometric algorithm for the
Maxwell system. (i) A lattice Maxwell system is developed with gauge
symmetry, symplectic structure, and discrete space-time symmetry.
(ii) Noether's theorem is generalized to the case of discrete symmetry
for the lattice Maxwell system, which establishes the correspondence
between discrete symmetries and discrete local conservation laws.
(iii) Applying the discrete Noether's theorem, the lattice Maxwell
system is shown to admit a discrete local energy-momentum conservation
law.

\appendix

\section{Discrete difference operators and identities}

In this appendix, we list the definitions of discrete operators and
discrete vector analysis identities that are needed in the present
study. 

The following are definitions of discrete operators. 
\begin{eqnarray}
\left({\nabla_{\mathrm{d}}}\phi\right)_{i,j,k} & = & [\phi_{i+1,j,k}-\phi_{i,j,k},\phi_{i,j+1,k}-\phi_{i,j,k},\phi_{i,j,k+1}-\phi_{i,j,k}]~,\label{EqnDEFGRADD}\\
\left(\CURLD\bfA\right)_{i,j,k} & = & \left[\begin{array}{c}
\left(A_{z,i,j+1,k}-A_{z,i,j,k}\right)-\left(A_{y,i,j,k+1}-A_{y,i,j,k}\right)\\
\left(A_{x,i,j,k+1}-A_{x,i,j,k}\right)-\left(A_{z,i+1,j,k}-A_{z,i,j,k}\right)\\
\left(A_{y,i+1,j,k}-A_{y,i,j,k}\right)-\left(A_{x,i,j+1,k}-A_{x,i,j,k}\right)
\end{array}\right]^{T}~,\label{EqnDEFCURLD}\\
\left({\DIVD}\bfB\right)_{i,j,k} & = & \left(B_{x,i+1,j,k}-B_{x,i,j,k}\right)+\left(B_{y,i,j+1,k}-B_{y,i,j,k}\right)\nonumber \\
 &  & \hspace{2cm}+\ \left(B_{z,i,j,k+1}-B_{z,i,j,k}\right)~,\label{EqnDEFDIVD}
\end{eqnarray}
\begin{eqnarray}
\left({\nabla_{\mathrm{d}}^{*}}\phi\right)_{i,j,k} & = & [\phi_{i,j,k}-\phi_{i-1,j,k},\phi_{i,j,k}-\phi_{i,j-1,k},\phi_{i,j,k}-\phi_{i,j,k-1}]~,\label{EqnDEFGRADD1}\\
\left(\CURLD^{T}\bfA\right)_{i,j,k} & = & \left[\begin{array}{c}
\left(A_{z,i,j,k}-A_{z,i,j-1,k}\right)-\left(A_{y,i,j,k}-A_{y,i,j,k-1}\right)\\
\left(A_{x,i,j,k}-A_{x,i,j,k-1}\right)-\left(A_{z,i,j,k}-A_{z,i-1,j,k}\right)\\
\left(A_{y,i,j,k}-A_{y,i-1,j,k}\right)-\left(A_{x,i,j,k}-A_{x,i,j-1,k}\right)
\end{array}\right]^{T}~,\label{EqnDEFCURLD1}\\
\left({\DIVD^{*}}\bfB\right)_{i,j,k} & = & \left(B_{x,i,j,k}-B_{x,i-1,j,k}\right)+\left(B_{y,i,j,k}-B_{y,i,j-1,k}\right)\nonumber \\
 &  & \hspace{2cm}+\ \left(B_{z,i,j,k}-B_{z,i,j,k-1}\right)~,\label{EqnDEFDIVD1}
\end{eqnarray}
\begin{eqnarray}
(\bfM*\times\bfB)_{i,j,k} & = & \left[\begin{array}{c}
M_{y,i+1,j,k}B_{z,i,j,k}-M_{z,i+1,j,k}B_{y,i,j,k}\\
M_{z,i,j+1,k}B_{x,i,j,k}-M_{x,i,j+1,k}B_{z,i,j,k}\\
M_{x,i,j,k+1}B_{y,i,j,k}-M_{y,i,j,k+1}B_{x,i,j,k}
\end{array}\right]^{T}~,\\
(\bfM\times*\bfB)_{i,j,k} & = & \left[\begin{array}{c}
M_{y,i,j,k}B_{z,i-1,j,k}-M_{z,i,j,k}B_{y,i-1,j,k}\\
M_{z,i,j,k}B_{x,i,j-1,k}-M_{x,i,j,k}B_{z,i,j-1,k}\\
M_{x,i,j,k}B_{y,i,j,k-1}-M_{y,i,j,k}B_{x,i,j,k-1}
\end{array}\right]^{T}~,\\
(\bfM\times_{2}\bfB)_{i,j,k} & = & \left[\begin{array}{c}
M_{y,i,j,k}B_{z,i,j,k}-M_{z,i,j,k}B_{y,i,j,k}\\
M_{z,i-1,j+1,k}B_{x,i,j,k}-M_{x,i-1,j+1,k}B_{z,i,j,k}\\
M_{x,i-1,j,k+1}B_{y,i-1,j,k}-M_{y,i-1,j,k+1}B_{x,i,j,k}
\end{array}\right]^{T}~,
\end{eqnarray}
\begin{eqnarray}
\left(\bfM\otimes Q_{x}\right)_{i,j,k} & = & [M_{x,i,j,k}Q_{x,i,j,k},M_{y,i,j,k}Q_{x,i-1,j+1,k},M_{z,i,j,k}Q_{x,i-1,j,k+1}]~,\\
\bfM_{J}\otimes_{2}Q_{x,J} & = & [M_{x,J}Q_{x,i-1,j+1,k+1},M_{y,J}Q_{x,i,j,k+1},M_{z,J}Q_{x,i,j+1,k}]~,\\
\left(\bfM\otimes^{*}Q_{x}\right)_{i,j,k} & = & [{M_{x}}_{i,j,k}{Q_{x}}_{i,j,k},{M_{y}}_{i-1,j+1,k}{Q_{x}}_{i,j,k},{M_{z}}_{i-1,j,k+1}{Q_{x}}_{i,j,k}]~.\\
\left(*\bfQ\right)_{i,j,k} & = & [Q_{x,i-1,j,k},Q_{y,i,j-1,k},Q_{z,i,j,k-1}]~.
\end{eqnarray}
The following are discrete vector analysis identities used in the
paper. For any discrete vector fields $\bfM_{J}$ and $\bfQ_{J}$,
\begin{eqnarray}
\CURLD\nabla_{d}M_{x,J} & = & 0~,\label{eq:63}\\
\DIVD\CURLD\bfM_{J} & = & 0~,\label{eq:64}\\
\CURLDP{\nabla_{d}}^{*}M_{x,J} & = & 0~,\label{EqnDIVDPCURLDP}\\
\DIVD^{*}\CURLDP\bfM_{J} & = & 0~,\label{eq:66}
\end{eqnarray}
\begin{eqnarray}
\bfM_{J}\cdot\CURLD\bfQ_{J}-\left(\CURLDP\bfM\right)\cdot\bfQ_{J}-\DIVD^{*}\left(\bfM_{J}*\times\bfQ_{J}\right) & = & 0~,\label{EqnCURLCUR2LDIV}\\
\left(\bfM\times*\CURLD\bfQ\right)_{J}-\left(\nabla_{d}^{*}\bfQ\right)_{J}\cdot\bfM_{J}+\bfM_{J}\cdot\left(\nabla_{d}*\bfQ\right)_{J} & = & 0~,\label{EqnCURLXDOTDOT}\\
\left(\DIVD^{*}\left(\bfM\otimes Q_{x}\right)\right)_{J}-{\left(\bfM_{J}\cdot\left(\nabla_{d}*\bfQ\right)\right)_{x}}_{J}-\left(\DIVD^{*}\bfM\right)_{J}{(*\bfQ)_{x}}_{J} & = & 0~,\label{EqnDIVDDOTDIRMUL}\\
\CURLDP\bfM_{J}\otimes Q_{x,J}-\CURLDP\left(\bfM_{J}\otimes_{2}Q_{x,J}\right)+\nabla_{d}Q_{x,J}\times_{2}\bfM_{J} & = & 0~,\label{EqnOTOT2CT2}\\
\left(\CURLD\bfQ_{J}\right)\otimes^{*}M_{x,J}-[\left(\CURLD\bfQ_{J}\right)\cdot\bfM_{J},0,0]+\nonumber \\
\left(\bfQ_{J}-\bfQ_{J-x}\right)*\times\bfM_{J}-\left(\nabla_{d}Q_{x,J}\right)\times_{2}\bfM_{J} & = & 0~.\label{EqnOTXDOTCTXCT2}
\end{eqnarray}

We note that in the continuous limit Eqs.\,\eqref{eq:63} and \eqref{EqnDIVDPCURLDP}
both recover the familiar identity $\nabla\times\nabla f=0$ for a
scalar field $f,$ and Eqs.\,\eqref{eq:64} and \,\eqref{eq:66}
both recover the identity $\nabla\cdot\nabla\times\bfM=0$ for a vector
field $\bfM.$ The structures embedded in these discrete vector analysis
identities are rich. A general discussion on this topic is beyond
the scope of the paper. However, we give a hint on these structures
by demonstrating that Eqs.\,(\ref{eq:63})-(\ref{eq:66}) can be
derived from two more general identities. Define discrete operators
$D_{\mathbf{a}}^{1}$, $D_{\mathbf{a}}^{2}$ and $D_{\mathbf{a}}^{3}$
as follows,

\begin{eqnarray}
D_{\mathbf{a}}^{1}\phi(x,y,z) & = & [\phi(x+a_{x},y,z)-\phi(x,y,z),\phi(x,y+a_{y},z)\nonumber \\
 &  & -\phi(x,y,z),\phi(x,y,z+a_{z})-\phi(x,y,z)],\\
D_{\mathbf{a}}^{2}\bfA(x,y,z) & = & \left[\begin{array}{c}
(A_{z}(x,y+a_{y},z)-A_{z}(x,y,z))-(A_{y}(x,y,z+a_{z})-A_{y}(x,y,z)),\\
(A_{x}(x,y,z+a_{z})-A_{x}(x,y,z))-(A_{z}(x+a_{x},y,z)-A_{z}(x,y,z)),\\
(A_{y}(x+a_{x},y,z)-A_{y}(x,y,z))-(A_{x}(x,y+a_{y},z)-A_{x}(x,y,z))
\end{array}\right],\label{eq:Da2}\\
D_{\mathbf{a}}^{3}\bfB(x,y,z) & = & \left(B_{x}(x+a_{x},y,z)-B_{x}(x,y,z)\right)+\left(B_{x}(x,y+a_{y},z)-B_{x}(x,y,z)\right)+\nonumber \\
 &  & \left(B_{z}(x,y,z+a_{y})-B_{z}(x,y,z)\right),\label{eq:Da3}
\end{eqnarray}
where $\phi$, $\bfA$ and $\bfB$ are any scalar and vector fields
on $R^{3}$, and $\mathbf{a}=[a_{x},a_{y},a_{z}]$ is an arbitrary
vector in $R^{3}$. It can be proven that $D_{\mathbf{a}}^{1}$, $D_{\mathbf{a}}^{2}$
and $D_{\mathbf{a}}^{3}$ satisfy the following identities,

\begin{eqnarray}
D_{\mathbf{a}}^{2}D_{\mathbf{a}}^{1}\phi(x,y,z) & = & 0,\\
D_{\mathbf{a}}^{3}D_{\mathbf{a}}^{2}\bfA(x,y,z) & = & 0.
\end{eqnarray}
If we let $\mathbf{a}=[1,1,1]$ and $\mathbf{a}=[-1,-1,-1]$, Eqs.
(\ref{eq:63})-(\ref{eq:66}) are immediately recovered.
\begin{acknowledgments}
This research is supported by National Magnetic Confinement Fusion
Energy Research Project (2015GB111003, 2014GB124005), National Natural
Science Foundation of China (NSFC-11575185, 11575186, 11305171, 11775219,
11775222), JSPS-NRF-NSFC A3 Foresight Program (NSFC-11261140328),
Chinese Scholar Council (201506340103), Key Research Program of Frontier
Sciences CAS (QYZDB-SSW-SYS004), the GeoAlgorithmic Plasma Simulator
(GAPS) Project, and U.S. Department of Energy (DE-AC02- 09CH111466).
\end{acknowledgments}

\bibliographystyle{apsrev4-1}
%

\end{document}